# Superconductivity emerging from suppressed large magnetoresistant state in WTe$_2$

Defen Kang[1]*, Yazhou Zhou[1]*, Wei Yi[1], Chongli Yang [1], Jing Guo[1], Youguo Shi[1], Shan Zhang[1], Zhe Wang[1], Chao Zhang[1], Sheng Jiang[2], Aiguo Li[2], Ke Yang[2], Qi Wu[1], Guangming Zhang[3,4], Liling Sun[1,4]†, Zhongxian Zhao[1,4]†

[1]Institute of Physics and Beijing National Laboratory for Condensed Matter Physics, Chinese Academy of Sciences, Beijing 100190, China

[2]Shanghai Synchrotron Radiation Facilities, Shanghai Institute of Applied Physics, Chinese Academy of Sciences, Shanghai 201204, China

[3] State Key Laboratory for Low dimensional Quantum Physics and Department of Physics, Tsinghua University, Beijing 100084, China

[4]Collaborative Innovation Center of Quantum Matter, Beijing, 100190, China

The recent discovery of large magnetoresistance (LMR) in WTe$_2$ provides a unique playground to find new phenomena and significant perspective for potential applications. The LMR effect originates from a perfect balance of hole and electron carriers, which is sensitive to external pressure. Here we report the suppression of the LMR and emergence of superconductivity in pressurized WTe$_2$ via high-pressure synchrotron X-ray diffraction, electrical resistance, magnetoresistance, and *ac* magnetic susceptibility measurements. Upon increasing pressure, the positive LMR effect can be gradually suppressed without crystal structure change and turned off at a critical pressure of 10.5 GPa, where superconductivity emerges unexpectedly. The superconducting transition temperature reaches to 6.5 K at ~ 13 GPa and then decreases to 2.6 K at ~ 24 GPa. *In-situ* high pressure Hall coefficient measurements at low temperature demonstrate that elevating pressure decreases hole carriers but increases electron ones. Significantly, at the critical pressure, a sign change in the Hall coefficient is observed, indicating a quantum phase transition of the Fermi surface reconstruction.

Tungsten ditelluride ($WTe_2$) is a well known non-magnetically thermoelectric semimetal[1-3]. Recently, its unexpected property of large magnetoresistance (LMR) has been discovered[4], yielding a research hotspot in condensed matter physics and material science. The LMR effect found in $WTe_2$ and some other nonmagnetic compounds[4-7] is a peculiar transport property. At ambient pressure, $WTe_2$ crystallizes in a crystal structure of distorted $MoS_2$[8], and the distortion is induced by the tungsten chains that are arranged along the *a* axis of the orthorhombic unit cell. The electrical resistance along the *a* axis increases dramatically when magnetic field is applied perpendicularly to the dichalcogenide layers (along to *c* axis), resulting in the LMR effect. Theoretical and experimental investigations indicate that the LMR effect is resulted from the perfect balance between electron and hole Fermi pockets along the Γ-X direction in the Brillouin zone[4,9,10], different from that of giant and colossal magnetoresistance previously found in magnetic materials[11,12]. Due to the complicated band structure with multiple two-carrier Fermi pockets in $WTe_2$, the fine details of the electronic structure play a significant role in the LMR effect. It is well-known that pressure as an important control parameter can effectively tune lattice structures and the corresponding electronic states in a more systematic fashion, avoiding the complexity brought by chemical doping[10,13-17]. Therefore, the high pressure studies on $WTe_2$ are particularly important to explore novel phenomena and understand the physics behind.

We first characterize the structure of $WTe_2$ sample at ambient pressure. Figure 1

shows X-ray diffraction pattern of powdered sample which is ground from a few pieces of single crystals. As it can be seen, the Bragg peaks in the pattern can be well indexed by orthorhombic structure. To clarify whether there is a structure change in pressurized $WTe_2$, we perform *in-situ* high-pressure synchrotron X-ray diffraction measurements. The results shown in Supplementary Figure 1 indicate no first-order structure phase transition under pressure up to 20.12 GPa, except for an additional lattice distortion at pressure above 13 GPa.

Then we perform the electrical resistance measurement for the single crystal under *quasi-hydrostatic* pressure. Since $WTe_2$ has an anisotropic electron structure, the LMR effect was discovered along the tungsten chain direction (*a* axis)[4]. To reveal the pressure effect on the LMR state, we apply the magnetic field and current in the same manner as ambient pressure. Figure 2 shows the typical temperature dependence of electrical resistance measured at zero magnetic field for pressures ranging from 0.3 GPa to 24 GPa. In Fig.2a, the electrical resistance curve at 0.3 GPa shares the similar behavior to that measured at ambient pressure[4,10]. Upon increasing pressure below 13.0 GPa, the electrical resistance is suppressed in the whole temperature range, while it is enhanced above 13 GPa. Intriguingly, it drops abruptly at 2.8 K and 10.5 GPa. Such a drop becomes more pronounced at higher pressures, and the zero electrical resistance is achieved between 11 - 24 GPa (Fig.2b).

To confirm whether the zero electrical resistance state is superconducting or not, we carry out *in-situ* high pressure alternating-current (*ac*) susceptibility measurements.

The Meissner effect is observed at the selected pressures of 15 GPa and 18.3 GPa, respectively (Fig.2c). The onset temperatures of the diamagnetism are consistent with that of the electrical resistance drop (Fig.2b). Both electrical resistance and magnetic measurements coordinately confirm that pressure induces a superconducting transition in $WTe_2$. The superconducting transition temperature (Tc) can reach to 6.5 K at ~ 13 GPa and monotonically decreases down to 2.6 K at ~24 GPa. It is noteworthy that, when the maximum Tc appears at the pressure of 13 GPa, the pressure starts to enhance the electrical resistance in the whole temperature range and induces additional lattice distortion as found in the high pressure X-ray diffraction pattern (Supplementary Information).

In order to reveal how the LMR state evolves into the superconducting state, we investigate the temperature dependence of electrical resistance for different given pressures under different magnetic fields. We find that the positive LMR effect of $WTe_2$ is suppressed by applying pressure (Fig.3a-3b). At the critical pressure ~11 GPa and above, the positive magnetoresistance effect no longer exists (inset of Fig.3c and 3d), and superconductivity appears simultaneously. Notably, the resistance as a function of temperature shows full suppression of superconductivity at 13 GPa under 3 Tesla (Fig.3e), indicating that the upper critical magnetic field of the superconducting $WTe_2$ is below 3 Tesla.

Moreover, we can define a characteristic temperature $T^*_{ZF}$ as the turn-on temperature of the LMR effect at zero field (as indicated by an arrow in the inset of Fig.2a). Such a definition of $T^*_{ZF}$ is coincident with the temperature of the linear extrapolation of turn-on LMR temperatures under different magnetic fields. Then we summarize our experimental results in the pressure-temperature phase diagram (Fig.4a). There are two distinct regions in the diagram: the LMR state and the superconducting state. It is found that the characteristic temperature $T^*_{ZF}$ of the LMR state decreases with increasing pressure and vanishes at the critical pressure 10.5 GPa. where the superconductivity emerges. The value of Tc starts to increase up to a maximum at 13 GPa and then declines with further increasing pressure. These results clearly demonstrate that the pressure can effectively suppress the LMR state and induce superconductivity.

It is known that the perfect balance between hole and electron populations accounts for the ambient-pressure LMR effect[4,9]. To understand the suppression of the LMR effect under pressure, we conduct the *in-situ* high pressure Hall coefficient ($R_H$) measurements at low temperature. The setup of our high-pressure Hall measurements leads to the detected Hall coefficient combining signals from two directions of parallel and perpendicular to the tungsten chains (see Method). Thus, the obtained results involve contributions from both kinds of carries at the Fermi surface in the Brillouin zone. Below 10.5 GPa, the Hall coefficient is positive and decreases with elevating pressure. Our results indicate that pressure can enhance the populations of electron carriers, but depopulate the hole carriers. The balance is thus deviated from its perfect state, which attributes to the suppression of the LMR effect (Fig.4b). These observations are in good agreement with the reported Shubnikov-de-Hass oscillation measurements[10]. On further increasing pressure, significantly, we found that the Hall coefficient $R_H$ suffers from a sign change from the positive to the negative at the critical pressure 10.5 GPa, where the LMR state vanishes and superconductivity emerges (Fig.4a). Above the critical pressure, the Hall coefficient is in a small negative value. At the critical pressure, the second derivative of the Hall coefficient with respect to pressure shows a maximum, which is expected to become divergent in the zero temperature limit. The sign change in $R_H$ is an indication of a significant reconstruction of Fermi surface. We thus propose that a quantum phase transition occurs at the critical pressure, separating the LMR state and superconducting state. Since such a kind of quantum phase transition with the changes of the Fermi surface structure can be characterized by the Lifshitz phase transition[18], the emergence of superconductivity observed in the pressurized $WTe_2$ may be connected with this transition, reminiscent of what is seen in Fe-based superconductors[19]. The mechanism of superconductivity in $WTe_2$ deserves further investigations from both experimental and theoretical sides.

Methods

High quality single crystals of $WTe_2$ were grown by a flux method, as described in Ref.4. Pressure was generated by a diamond anvil cell (DAC) with two opposing anvils sitting on a Be-Cu supporting plate. Diamond anvils of 300 μm flats were used for this study. A nonmagnetic rhenium gasket with 100 μm diameter hole was used for different runs of the high-pressure studies. The four-probe method was applied in the *ab* plane of single crystal $WTe_2$ for all high pressure transport measurements. For the high-pressure Hall coefficient measurements, the van der Pauw method was applied in this study. A constant current goes through a diagonal of the squared sample and the Hall voltage is measured from the other diagonal. To keep the sample in a quasi-hydrostatic pressure environment, NaCl powder was employed as the pressure medium. The high pressure alternating-current susceptibilities were detected within a lock-in amplifier in a diamond-anvil cell, with a signal coil around the diamond anvils and a compensating coil[20]. High pressure X-ray diffraction (XRD) experiments were performed at beam line 15U at the Shanghai Synchrotron Radiation Facility. Diamonds with low birefringence were selected for the experiments. A monochromatic X-ray beam with a wavelength of 0.6199 Å was adopted for all XRD measurements. To maintain the sample in a hydrostatic pressure environment, silicon oil was used as pressure medium in the high-pressure XRD measurements. Pressure was determined by the ruby fluorescence method[21].

**Acknowledgements**

We thank V. Sidorov for helpful discussions. The work was supported by the NSF of China (Grant No. 91321207, 11427805), 973 projects (Grant No.2011CBA00100 and 2010CB923000) and the Strategic Priority Research Program (B) of the Chinese Academy of Sciences (Grant No. XDB07020300 and XDB07020100).

†To whom correspondence should be addressed.

E-mail: llsun@iphy.ac.cn and zhxzhao@iphy.ac.cn.

* These authors contributed equally to this work.

***Note added:*** when this paper is written, we note an arXiv paper (1501. 07379), which reports the evidence of pressure-induced superconductivity in $WTe_2$ from electrical resistance measurements without pressure transmitting medium.

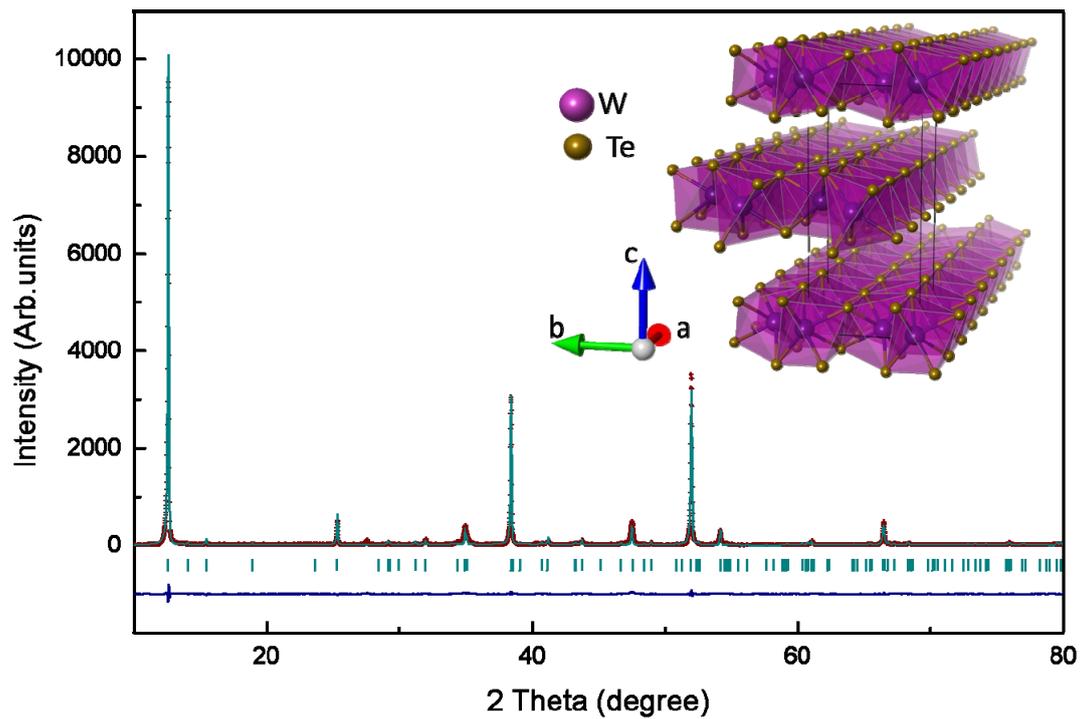

Figure 1. Ambient-pressure structure information of WTe$_2$ sample. The main figure shows our experimental and indexed X-ray diffraction patterns of WTe$_2$. The bars display the Bragg reflection positions. The insert exhibits its three dimensional layer structure constructed by edge-shared WTe$_6$ octahedron.

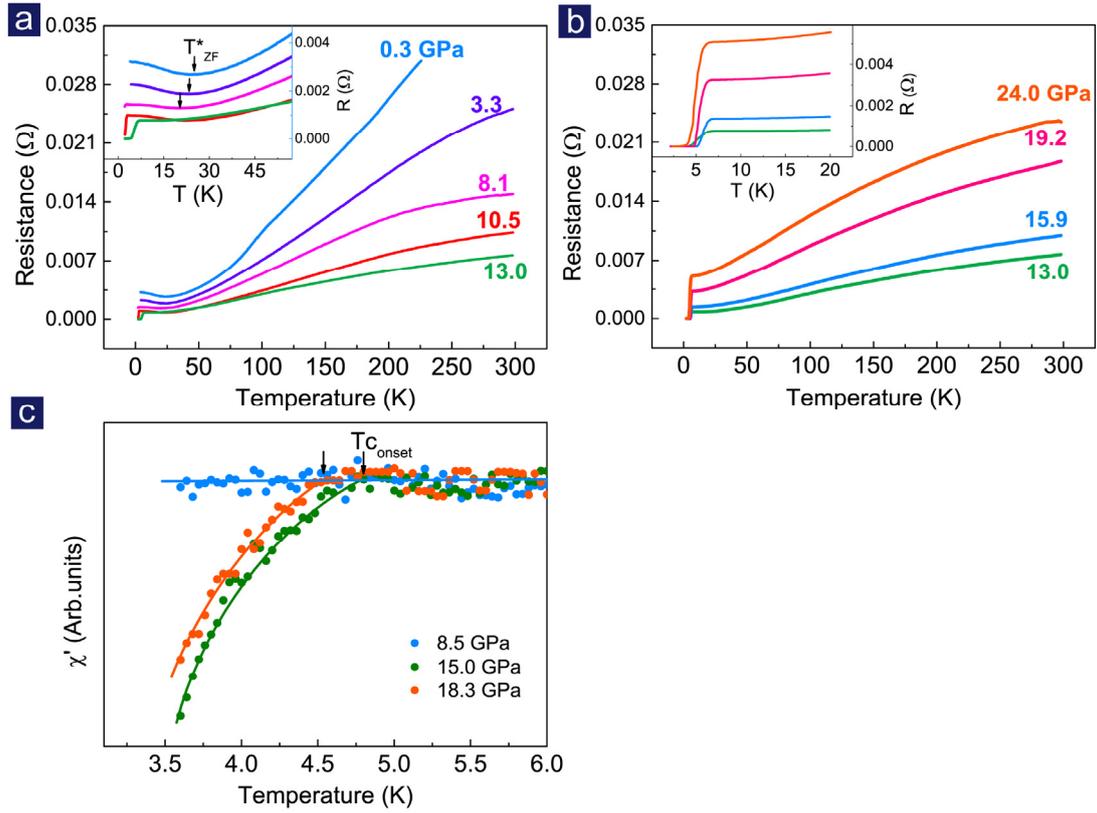

Figure 2 Temperature dependence of electrical resistance and alternating-current (*ac*) magnetic susceptibility for single crystal WTe₂ at different pressures. (a) The plot of electrical resistance as a function of temperature measured without magnetic field for the pressures ranging from 0.3 GPa to 13.0 GPa. The inset displays the upturn of electrical resistance in the low temperature range, where a characteristic temperature is defined by the minimum resistance and taken as turn-on temperature ($T^*_{ZF}$) of LMR effect, as indicated by arrows. (b) Temperature dependence of electrical resistance measured at different pressures between 13.0 GPa and 24 GPa. The inset shows clear electrical resistance drops and zero resistance behavior. (c) The real part of the *ac* magnetic susceptibility ($\chi'$) versus temperature for the single crystal WTe₂ at different pressures, and the diamagnetism is confirmed at the selected pressures of

15.0 GPa and 18.3 GPa, respectively.

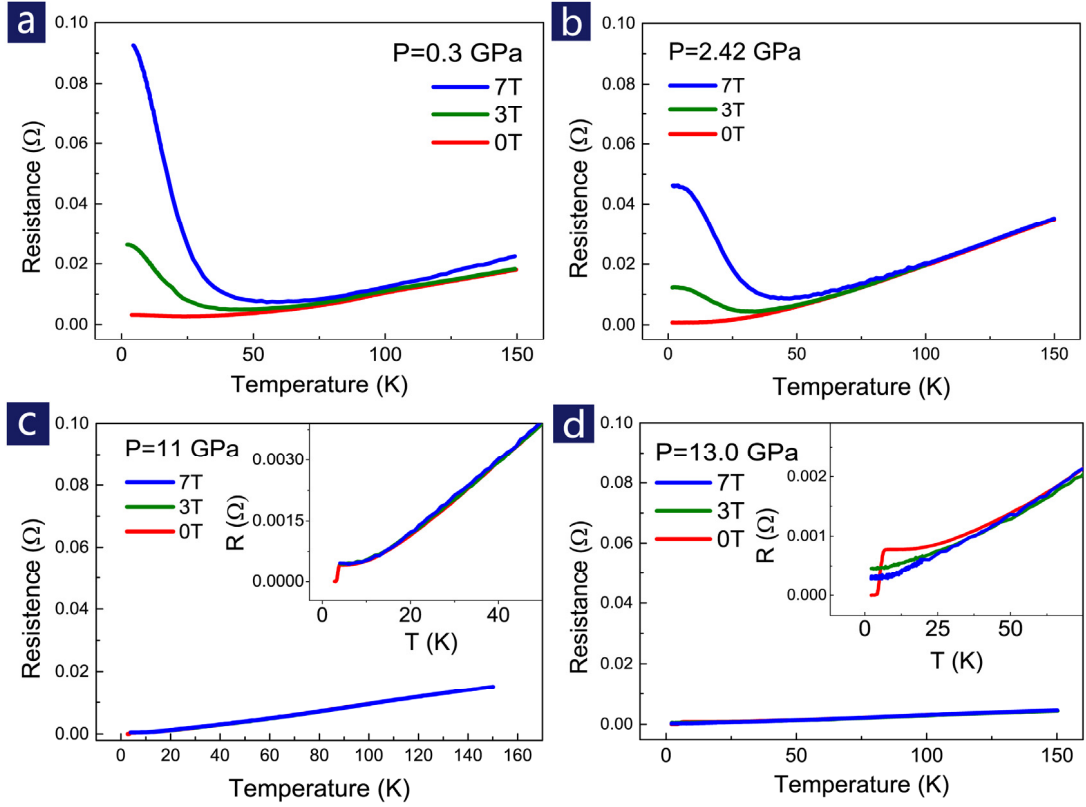

Figure 3 High-pressure resistance versus temperature under different magnetic fields. (a)-(b) The electrical resistance as a function of temperature at 0.3 GPa and 2.42 GPa, respectively, illustrating the dramatic suppression of the LMR effect. (c) The plot of resistance versus temperature at 11 GPa, showing a full suppression of the positive magnetoresistance and a zero resistance at 3.5 K. (d) The temperature dependence of electrical resistance at 13 GPa.

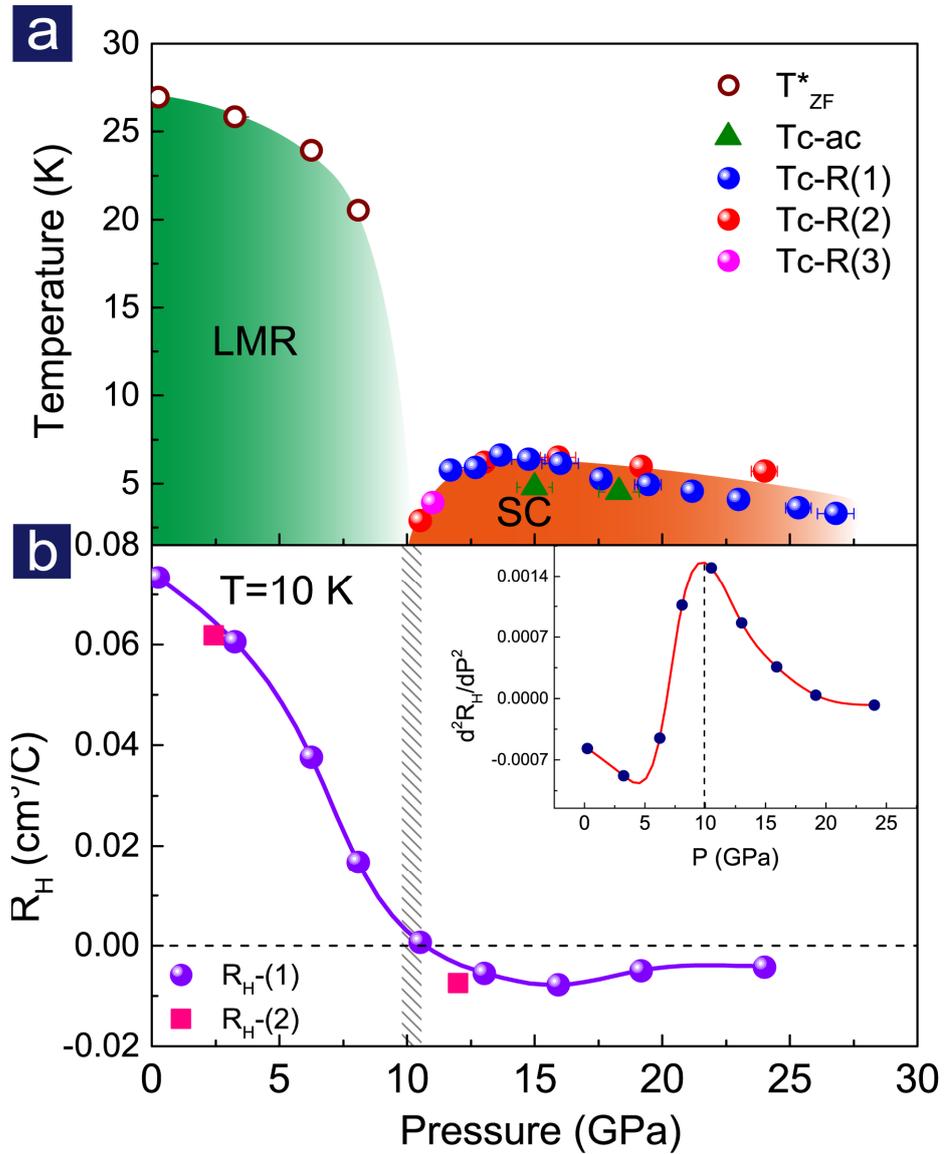

Figure 4    Pressure dependences of the characteristic temperatures and Hall coefficient of WTe₂. (a) The plot of T*₍ZF₎ and Tc versus pressure. The red, pink and blue solid circles represent Tc extracted from different runs of electrical resistance measurements, and the green triangles represent the Tc determined from the *ac* susceptibility measurements. The acronyms LMR and SC stand for large magnetoresistance and superconductivity, respectively. (b) Hall coefficient (R_H) as a function of pressure measured at 10 K and 1 Tesla, displaying a sign change from the positive to the negative at the critical pressure. Solid purple circles and pink squares represent the $R_H$ obtained from different runs. The inset shows the second derivative of the Hall coefficient and the maximum corresponds to the sign change of Hall coefficient.